\documentclass[useAMS,usenatbib,asm]{mn2e}
 \usepackage{graphicx,fleqn,times,amssymb,upgreek}

\title{Open-cluster density profiles derived using a kernel estimator}

\author[Anton F. Seleznev]{Anton F. Seleznev$^{1}$\thanks{E-mail:
anton.seleznev@urfu.ru (AFS)}\\
$^{1}$Astronomical Observatory, Ural Federal University, Mira str.
19, Ekaterinburg 620002, Russia}

\begin{document}

\date{Accepted 2015..... Received 2015...; in original form 2015...}

\pagerange{\pageref{firstpage}--\pageref{lastpage}} \pubyear{2015}

\maketitle

\label{firstpage}

\begin{abstract}
Surface and spatial radial density profiles in open clusters are derived
with the use of a kernel estimator method. Formulae are obtained for
contribution of every star into spatial density profile. Evaluation of spatial density
profiles is tested against open cluster models from N-body experiments with
N=500. Surface density profiles are derived for seven open clusters NGC 1502,
NGC 1960, NGC 2287, NGC 2516, NGC 2682, NGC 6819 and NGC 6939 by means of
2MASS data and for different limiting magnitudes. The selection of optimal kernel
halfwidth is discussed. It is shown that open cluster radius
estimates hardly depend on kernel halfwidth. Hints of stellar mass
segregation and structural features indicating cluster
non-stationarity in the regular force field are found. A comparison with other
investigations shows that the data
on open cluster sizes are often underestimated. The existence of an
extended corona around open cluster NGC 6939 was confirmed. A combined function composed of King density profile for the cluster
core and uniform sphere for the cluster corona is shown to be a better approximation of the surface radial density profile.
King function alone does not
reproduce surface density profiles of sample clusters properly.
Number of stars, the cluster masses, and the tidal radii in the Galactic gravitational
field for the sample clusters are estimated. It is shown that NGC 6819 and
NGC 6939 are extended beyond their tidal surfaces.
\end{abstract}


\section{Introduction}
Surface density profiles are traditional tools in investigations of the structure of stellar
clusters. Surface density profiles were used for cluster size
determination, for example, \citet*{SSB}, \citet{SSGH},
\citet*{CBB}. It can be noted
that usually surface density profiles were plotted as histograms of star counts,
and stochasticity of histograms prevented a reliable cluster size determination.
Methods were presented to reduce both stochasticity and asymmetries. Kholopov and Artyukhina performed
star counts in the series of overlapping rings of different widths and in
overlapping sectors, see, for example, \citet{AK_M67}, \citet{K_M15}. \citet{Dj} proposed
an averaging of star counts across several angular bins. Apart from stochasticity, the
limited field of view is often the reason for unreliability of cluster size determination.

Cluster density profiles can be compared with different dynamic models in order to reveal the results of different
dynamic processes. For example, gravothermal catastrophe in globular clusters
becomes apparent by means of post-collapse density profiles \citep{SK95, SK97, MLFDVPBPS}.
Density profiles in the outer cluster
parts reveal cluster disruption processes in the outer tidal field, for example, \citet*{CZMB},
\citet{KKBH}, \citet{CGSKMP}.
The presence of mass
segregation shows an efficiency of stellar encounters; or, in the case of
extremely young clusters, preferential birth places of stars with different
masses or special features in the cluster formation process: for example, \citet{PGAGAHMD}, \citet{Hyades}, \citet*{VMMPZ},
\citet{GBSH}. Irregularities in the density profiles
indicate non-stationarity of a cluster in the regular field
\citep{DP}.

The extended sparse outer regions of open star clusters, i.e. cluster coronae, are of special interest.
The modern review of arguments in favour of cluster coronae existence was presented
in \citet*{DPS}. The cluster coronae can extend over the open cluster tidal surface.
Stars leave the cluster through the tidal surface in the vicinity of Lagrange points
(see, for example, \citet*{KMH}, \citet{KKBH1}). Part of these stars goes fast at large
distances from the cluster and forms the cluster tidal tails. Another part of these stars,
before moving to tidal tails, can live in the close
cluster vicinity (up to distances of four tidal radii of the cluster in the Galactic gravitational
field) for a relatively long time, comparable with the mean lifetime of the
cluster \citep{DPS}. It is the cluster corona. The formation of coronae
in open clusters and in their numerical models can be explained by the formation of
unstable periodic orbits and the large number of retrograde unclosed trajectories in the
vicinity of such orbits \citep{DPS}.

The detection of the open cluster coronae is difficult due to low stellar density in the coronae,
and due to fluctuations of the stellar density of the background. The
parameters of the open cluster coronae can be determined
more firmly and reliably after identifying probable cluster members, taking into account the data on
the stellar proper motions, see, for example, \citet{A}. \citet*{DMP} proposed the method
of star counts (referred hereafter as DMP), based on the use of the function $N(r)$, the number of stars in the circle of
radius $r$. This method was used by \citet{DS94} for the study of the structure of 103 open
star clusters. The method implies the comparison of the cluster field with several fields of
the cluster neighbourhood. This requires the study of a very large region around
the cluster (with the radius of up to six cluster radii). The use of this method is
restricted by large-scale fluctuations of the stellar background density in the cluster
vicinity. The goal of the present paper is the use of surface density function $F(r)$,
derived with the kernel estimator, for the search of coronae of the open clusters.

Surface density $F(r)$ is the number of stars per unit area of the celestial
sphere ($r$ is the current distance from the cluster centre, $R$ is the radius of the circle (sphere)
around the cluster centre).

\begin{equation}
\label{surfdens}
dN=2\pi rF(r)dr\; \mbox{,}  \qquad
N=2\pi \int\limits_0^R F(r)rdr \; \mbox{.}
\end{equation}

Spatial density $f(r)$ is the number of stars per unit volume of the coordinate
space.

\begin{equation}
\label{spatdens}
dN=4\pi r^2f(r)dr\; \mbox{,}  \qquad
N=4\pi \int\limits_0^R f(r)r^2dr \; \mbox{.}
\end{equation}

The use of radial density profiles assumes the hypothesis of a spherical symmetry.
Both surface and spatial stellar density are connected with the corresponding
probability densities.

\begin{equation}
\label{surfdensprob}
\varphi(r)=\frac{2\pi r}{N}F(r)\; \mbox{,} \qquad
\int\limits_0^R \varphi(r)dr=1 \; \mbox{.}
\end{equation}

\begin{equation}
\label{spatdensprob}
\psi(r)=\frac{4\pi r^2}{N} f(r)\; \mbox{,}  \qquad
\int\limits_0^R \psi(r)dr=1 \; \mbox{.}
\end{equation}

Consequently, methods of probability density evaluation can be used to get
surface and spatial density. Such methods were considered by
\citet{Silverman}. The kernel estimator stands out among them by intuitive
clarity and relatively simple realization. The essence of the kernel estimator
method is the following: every data point in the sample is replaced by some
function (kernel) normalized by 1. The result of the probability density
is the sum of all kernels divided by the number of sample points $N$.
Estimates of the surface or spatial density are obtained as the sum
of kernels, not divided by $N$.
It is very important that the density estimate inherits the properties of the kernel
function; for example, continuity and differentiability in the case of kernels
used in this paper.

The kernel estimator was used in the previous research for estimates of luminosity function and for
deriving and analysing surface density maps in star clusters
\citep{Sel,SCPR,PCPSSS,S235,SCCL,CS}.

\citet{MT} used the kernel estimator and the maximum penalized
likelihood estimator for the estimation of density profiles. They showed that
the one-dimensional kernel estimator did not suit for a surface density profile
construction and a two-dimensional method was needed. \citet{MT}
obtained formulae for a kernel function for the case of the surface radial
density profile and got estimates for spatial density solving an Abel
equation. They investigated the efficiency of both methods for three important
distributions (Plummer, de Vaucouleurs, Michie-King) and showed that the use
of an `optimal' kernel halfwidth, determined with the minimization of the integrated
mean-square error, led to an unsatisfactory result. \citet{MT}
proposed an empirical selection of kernel halfwidths, namely getting a
series of profile estimates and selecting the best version, that is `simply looking
at plots produced using several different values of the smoothing parameter, and
accepting the one that is as smooth as possible without being obviously biased -- that
is, the smoothest curve that closely follows the mean trend defined by curves computed
with much smaller smoothing parameter'. They used both kernel and maximum penalised
likelihood methods for deriving surface density profiles for the Coma cluster of galaxies
and for the M15 globular cluster.

In the present work, a kernel estimator is used for constructing surface radial
density profiles for seven open clusters; and for constructing spatial radial
density profiles for the numerical models of the open cluster coronae obtained by
N-body experiments with $N=500$. The paper is organized as follows. Section 2
is devoted to the development of formulae for surface and spatial density profiles.
Spatial density profiles of coronae of the N-body open cluster models are derived in
Section 3. Section 4 contains the description of the surface density radial profiles
derivation for seven open clusters and the discussion of the profiles. The estimation
of the cluster sizes is discussed in Section 5, the results of the present paper
are compared with the data from literature. Section 6 describes an approximation
of the cluster radial surface density profiles by King profile, with and without
considering the contribution from the cluster corona. The cluster mass and
the tidal radii estimates are obtained in Section 7. Conclusions are given in
Section 8.

\section{Kernel estimator for surface and spatial radial density profiles}

To understand the derivation of the formulae better, let us begin with the case of the surface density profile.
Consider the plane $(x,y)$ tangent to the
celestial sphere at the point of cluster centre O (see Fig. 1). Point
S is a projection of a star to the tangent plane, a circle with centre S is the
projection of the kernel with halfwidth $h$; $r_{\ast}$ is the distance of the star
from the cluster centre in the projection.
The contribution of this star to the surface density profile estimate
at the distance $r_{\rm i}$ from the cluster centre is evaluated. The kernel $K_2$ (see Eq.(4.5),
\citet{Silverman}) is used for the calculation of the surface
density. This kernel corresponds to the contribution to the surface
density as:

\begin{equation}
\label{kernel1}
\Delta F = \left\{
	    \begin{array}{ll}
	    \frac{\displaystyle 3}{\displaystyle \pi h^2}\left(1-\frac{\displaystyle \rho^2}
             {\displaystyle h^2}\right)^2 & \mbox{with}\quad \rho<h \; \mbox{,}\\
	    0 & \mbox{with}\quad  \rho \ge h  \; \mbox{.}\\
	    \end{array}
	   \right.
\end{equation}

This kernel function (often named as `quartic' kernel) has an advantage in the computational aspect.
Namely, this function has high smoothness properties contrary to Epanechnikov kernel, that allow to use a reasonably
coarse grid for contouring without introducing appreciable errors \citep{Silverman}, it is important
especially when plotting two-dimensional maps of surface density. Another one,
Gaussian kernel, is excellent in differentiability, but it requires much greater amount of computations
\citep{MT}.

\begin{figure}
   \centering
   \includegraphics[width=8truecm]{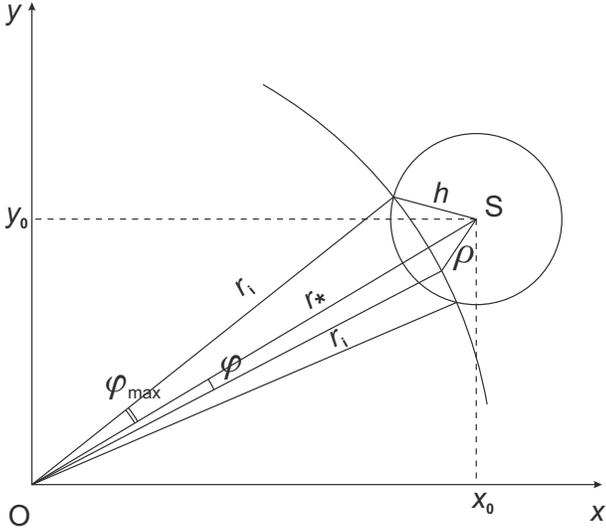}
   \caption{The plane $(x,y)$ is the tangent plane to celestial sphere at the point of the cluster centre O.
    Point S is a projection of a star to the tangent plane, a circle with centre S is the
    projection of the kernel with halfwidth $h$; $r_{\ast}$ is the distance of the star
    from the cluster centre in the projection. The case $|r_\ast -r_{\rm i}|< h$.}
   \label{fig1}
   \end{figure}

In order to get the contribution of star S into the surface density profile
at the distance $r_{\rm i}$ from the cluster centre, we need
to integrate this function by $\varphi$ over the arc of the circle with radius $r_{\rm i}$ from
$-\varphi_{\rm max}$ to $\varphi_{\rm max}$ (it is the case when $|r_\ast -r_{\rm i}|< h$) (see Fig.1).
The result is:

\begin{equation}
\label{surfdens1}
\begin{array}{l}
\Delta F(r_{\rm i})=\frac{\displaystyle 3}{\displaystyle \pi^2 h^2}\left(1-\frac{\displaystyle r_{\rm i}^2+r_\ast^2}{\displaystyle h^2}\right)^2\varphi_{\rm max}
	      +\frac{\displaystyle 6r_{\rm i}^2r_\ast^2}{\displaystyle \pi^2 h^6}\varphi_{\rm max} \\
\\
{} +\frac{\displaystyle 12r_{\rm i} r_\ast}{\displaystyle \pi^2 h^4}\left(1-\frac{\displaystyle r_{\rm i}^2+r_\ast^2}{\displaystyle h^2}\right)\sin\varphi_{\rm max}
              +\frac{\displaystyle 3r_{\rm i}^2r_\ast^2}{\displaystyle \pi^2 h^6}\sin2\varphi_{\rm max} \; \mbox{,}
\end{array}
\end{equation}

\noindent where

$$
\varphi_{\rm max}=\cos^{-1}\left(\frac{r_{\rm i}^2+r_\ast^2-h^2}{2r_{\rm i} r_\ast}\right) \; \mbox{.}
$$

Another situation is possible: when the circle of radius $r_{\rm i}$ lies inside
the circle of the kernel (see Fig.2, $r_{\rm i}< h-r_\ast$). In this case we need to
integrate Eq.(5) by $\varphi$  from $0$ to $2\pi$. The result is:

\begin{equation}
\label{surfdens2}
\Delta F(r_{\rm i})=\frac{3}{\pi h^2}\left(1-\frac{r_{\rm i}^2+r_\ast^2}{h^2}\right)^2
	      +\frac{6r_{\rm i}^2r_\ast^2}{\pi h^6} \; \mbox{.}
\end{equation}

It is easy to show that Eq.(6) and Eq.(7) coincide with Eq.(28b) from \citet{MT}.

\begin{figure}
   \centering
   \includegraphics[width=8truecm]{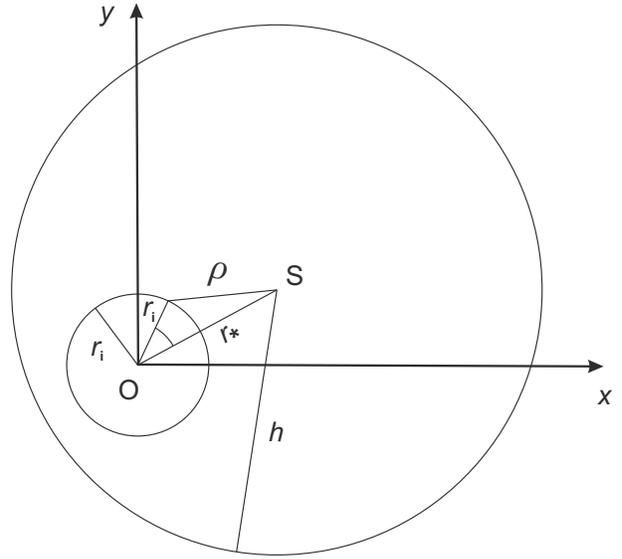}
   \caption{The same as in the Fig.1, but for the case $r_{\rm i}< h-r_\ast$.}
   \label{fig2}
   \end{figure}

The same approach is used for the determination of the contribution of the star
into spatial density when the spatial coordinates $(x,y,z)$ of the star are known.
The multivariate Epanechnikov kernel
(see Eq.(4.4), \citet{Silverman}) for three dimensions is used for the case of spatial density. It corresponds
to the contribution to spatial density as:

\begin{equation}
\label{kernel2}
\Delta f = \left\{
	    \begin{array}{ll}
	    \frac{\displaystyle 15}{\displaystyle 8\pi h^3}\left(1-\frac{\displaystyle \rho^2}{\displaystyle h^2}\right) & \mbox{with}\quad \rho<h \; \mbox{,}\\
	    0 & \mbox{with}\quad \rho \ge h \; \mbox{.}\\
	    \end{array}
	   \right.
\end{equation}

The Epanechnikov kernel in the case of three dimensions was selected also due to
the computational consideraton. It gives more simple equations for the density profile
in contrast to the quartic kernel, and requires less number of computations
in contrast to the Gaussian kernel. In addition, the difference between the Epanechnikov,
quartic, and Gaussian kernels is very little in many aspects \citep{Silverman,MT}.

Fig.3 shows star S at distance $r_\ast$ from cluster centre O
and three-dimensional kernel with the halfwidth $h$. The contribution
of this star to the spatial density profile at distance $r_{\rm i}$ from the cluster
centre is calculated. Fig.4 shows the sphere with radius $r_{\rm i}$ around the cluster centre.
The coordinate system in Fig.4 was transformed into $(\xi,\eta,\zeta)$ with
axis $\zeta$ in the direction from the cluster centre to star S.
In order to get the required contribution, it is necessary to integrate function Eq.(8)
over the segment of this sphere
by $\theta$ from $0$ to $2\pi$ and by $\varphi$ from $0$ to $\varphi_{\rm max}$
in the case shown in Fig.4 ($|r_\ast -r_{\rm i}|< h$) or from $0$ to $\pi$ in the case
when the sphere of radius $r_{\rm i}$ lies inside the sphere of kernel ($r_{\rm i}<
h-r_\ast$). The result is the following. For the case $|r_\ast -r_{\rm i}|< h$:

\begin{equation}
\label{spatdens1}
\begin{array}{l}
\Delta f(r_{\rm i})=\frac{\displaystyle 15}{\displaystyle 16\pi h^3}\left(1-\frac{\displaystyle r_{\rm i}^2+r_\ast^2}{\displaystyle h^2}\right)
               \left(1-\cos\varphi_{\rm max}\right) \\
               \\
	\qquad \qquad {}       +\frac{\displaystyle 15r_{\rm i}r_\ast}{\displaystyle 32\pi h^5}\left(1-\cos2\varphi_{\rm max}\right) \; \mbox{,}
\end{array}
\end{equation}

\noindent where $\varphi_{\rm max}$ is defined just as in Eq.(6). And for the case
$r_{\rm i}<h-r_\ast$:

\begin{equation}
\label{spatdens2}
\Delta f(r_{\rm i})=\frac{15}{8\pi h^3}\left(1-\frac{r_{\rm i}^2+r_\ast^2}{h^2}\right) \; \mbox{.}
\end{equation}

\begin{figure}
   \centering
   \includegraphics[width=8truecm]{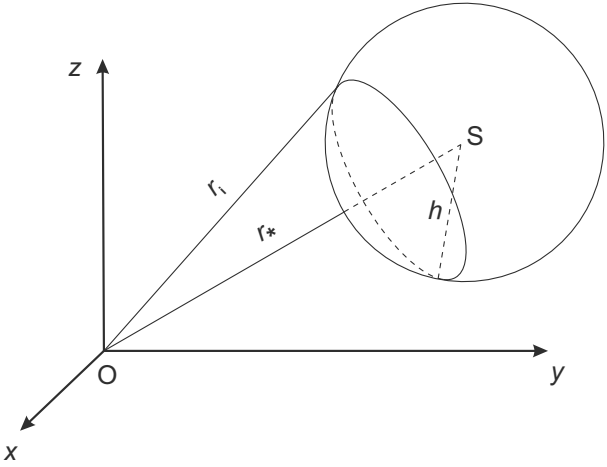}
   \caption{Star S at distance $r_\ast$ from cluster centre O
and three-dimensional kernel with halfwidth $h$. The case of $|r_\ast -r_{\rm i}|< h$.}
   \label{fig3}
   \end{figure}

\begin{figure}
   \centering
   \includegraphics[width=8truecm]{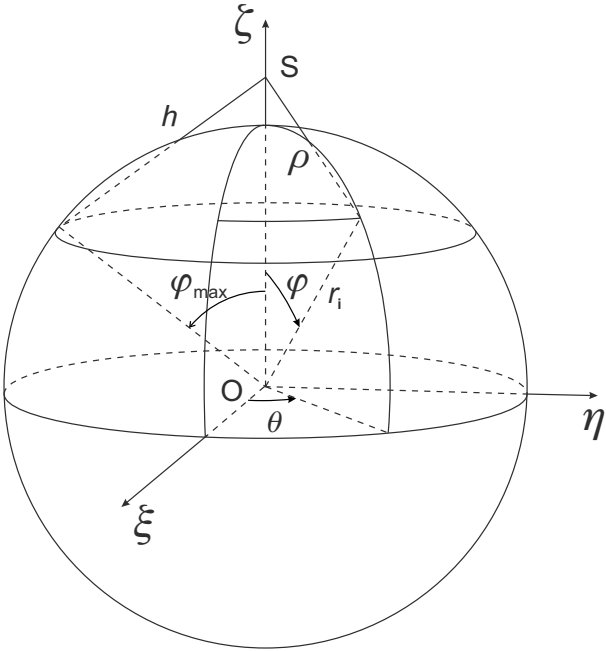}
   \caption{The sphere with radius $r_{\rm i}$ around cluster centre O. The case $|r_\ast -r_{\rm i}|< h$.}
   \label{fig4}
   \end{figure}

The algorithm of estimating both spatial and surface density is simple. One
must go over the sample of stars, determine at what numbers $i$ (distances
$r_{\rm i}$) every star contributes to the density and sum up these contributions
in accordance with the formulae listed above into array cells with numbers $i$.
Both fixed and adaptive kernel estimator algorithms were examined in the present paper
\citep{Silverman,MT}. An idea of the adaptive kernel algorithm consists in the use
of the kernels with different halfwidths depending on the density value. The adaptive
kernel estimator gives better estimates in the wings of distribution \citep{Silverman}.
This algorithm takes two steps: at the first step the pilot density estimate
is obtained with the fixed kernel algorithm; this pilot estimate is used at the second
step for determination of the kernel halfwidth through factors $\lambda$. The adaptive
kernel algorithm is described in details in \citet{Silverman}, and \citet{MT}.
In the present paper the same kernel function is used at both steps.

\section{Spatial density profiles of coronae of N-body open cluster models}

At present, the information about the spatial coordinates of stars in star
clusters is not available. In order to derive a spatial radial density profile, one must use
methods like Zeipel's or Plummer's; or solve the Abel equation numerically. All
these methods require making assumptions about the symmetry type. But the
situation will change when GAIA data are available. These data
will allow the study of cluster spatial structures directly, at least for the nearest
star clusters. Indeed, parallaxes from GAIA data will have standard errors
$(5-14)\; \upmu {\rm as}$ for stars in the magnitude range of $V\in (6,12)\; \rm{mag}$
and $(9-26)\; \upmu {\rm as}$ for stars with $V=15\; \rm{mag}$
\citep{GAIA}. In the case of Pleiades cluster with the distance of 120.2 pc \citep{Pl_dist}
it gives a distance error in the limits of 0.2 pc for bright stars, and of 0.4 pc
for stars with $V=15\; \rm{mag}$. With the linear radius of Pleiades
of about 10 pc \citep{Pl_rad}, this accuracy is sufficient for the study of the spatial structure
of this cluster. Pleiades have about a hundred of stars in the magnitude range of
$V\in (6,12)\; \rm{mag}$ \citep{Pl_LF}.

In the present paper the use of a kernel estimator
for the construction of spatial density profiles is illustrated, with spatial coordinates of
stars obtained by N-body simulations.

The kernel estimator was used previously for deriving surface radial density profiles
of open cluster corona models obtained by numerical N-body experiments, with
N=500 \citep{DD}. It was found that the stars, leaving the cluster and forming the cluster
corona, shape the surface density distribution close to equilibrium at the
distances from the cluster centre in the range from one to three cluster tidal radii \citep{DPS}.

Spatial radial density profiles were derived in the present work
with the use of Eq.(9) and Eq.(10) for the same N-body model outputs.
The adaptive kernel algorithm was used, because the outer part of the cluster model
corona has a very low density. Selection of the optimal kernel halfwidth was made
following the recommendations of \citet{MT}. Fig.5 shows the spatial density
profiles of the open cluster corona model 1 from \citet{DD} at the time point of about 150 Myrs
(about three violent relaxation times of the model), obtained with the
different kernel halfwidths (0.5, 1, 2, 3, 4, and 5 pc from top to bottom).
The halfwidth mentioned everywhere in this section is the one used in the pilot estimate
for the adaptive kernel method. The stochasticity
of the plots in the central region of the cluster is caused by small values of
factors $\lambda$, that control the kernel halfwidth in the adaptive kernel algorithm
($\lambda<1$ for $r<10$ pc). Due to this reason, factors $\lambda$ were restricted
in the present work by 1 from the lower side in the case of spatial density determination.

\begin{figure}
   \centering
   \includegraphics[width=8truecm]{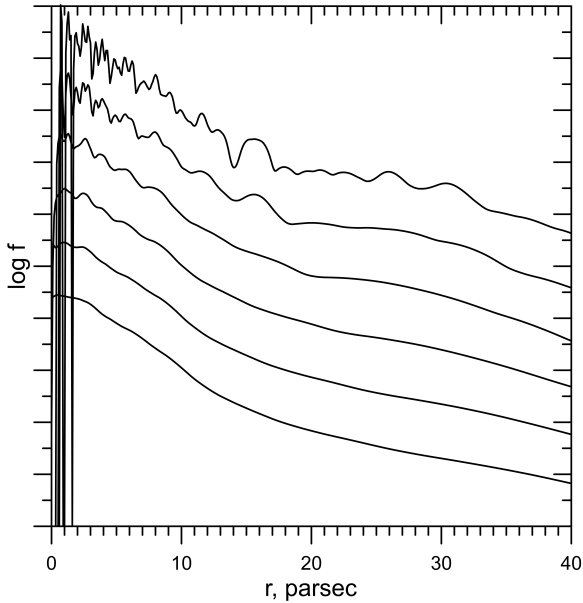}
   \caption{Spatial radial density profiles for corona of model 1 from Danilov \& Dorogavtseva (2008),
   the time point of about 150 Myrs. The kernel halfwidths are 0.5, 1, 2, 3, 4, and 5 pc from top to bottom.
   The vertical axis shows the logarithm of the spatial density (the density units
   are $\rm pc^{-3}$). The major ticks at the vertical axis differ by 1 dex, plots
   are shifted from each other by the value of 1 dex. The horizontal axis shows
   the distance from the cluster centre in parsecs.}
   \label{fig5}
   \end{figure}

\begin{figure}
   \centering
   \includegraphics[width=8truecm]{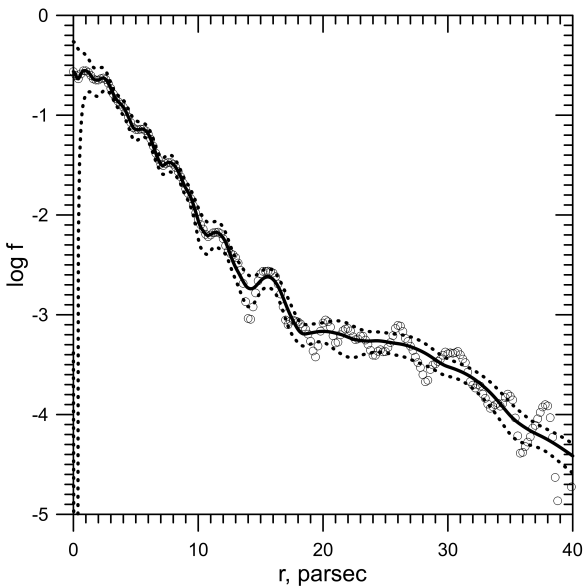}
   \caption{Comparison of the adaptive and fixed kernel estimates of spatial density
   of the open cluster corona model. The solid line is the adaptive estimate,
   the dotted lines show the confidence interval of $2\sigma$ width, open circles
   show the fixed kernel estimate. The kernel halfwidth is 1 pc. In the case of
   adaptive kernel estimator it is the kernel halfwidth for the pilot estimate.
   The time point is about 150 Myrs.}
   \label{fig6}
   \end{figure}

\begin{figure}
   \centering
   \includegraphics[width=8truecm]{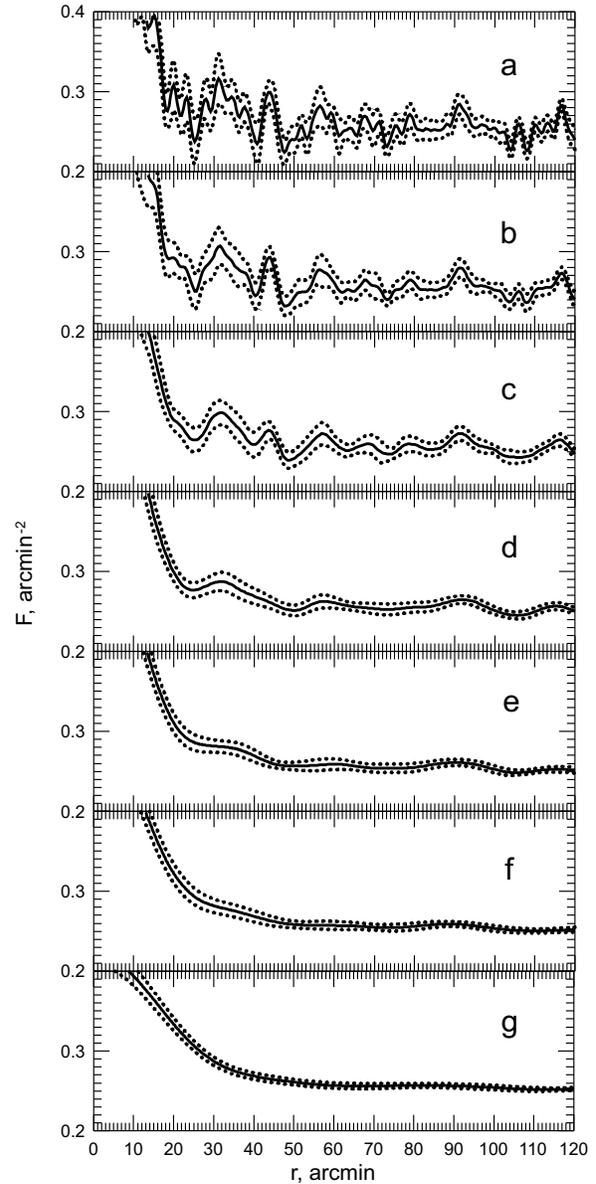}
   \caption{Surface density profiles of open cluster NGC 2287, obtained with different kernel
   halfwidth values for $J_{\rm lim}=13$ mag. (a) $h=2$ arcmin, (b) $h=3$ arcmin, (c) $h=5$ arcmin,
   (d) $h=10$ arcmin, (e) $h=15$ arcmin, (f) $h=20$ arcmin, (g) $h=30$ arcmin.
   The ordinate is the surface density in the units of $\rm arcmin^{-2}$, the abscissa is the distance
   from the cluster centre in arcmin.
   The thick solid line shows the surface density kernel estimate and the dotted lines show
   the confidence interval of $2\sigma$ width, obtained by a smoothed bootstrap method.}
   \label{fig7}
   \end{figure}

\begin{figure}
   \centering
   \includegraphics[width=8truecm]{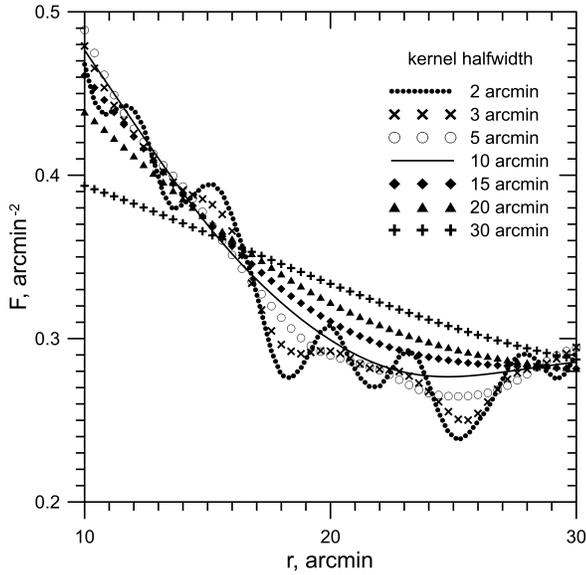}
   \caption{Surface density profiles of open cluster NGC 2287, obtained with different kernel
   halfwidth values for $J_{\rm lim}=13$ mag in the transition region between the cluster core
   and the cluster halo.  The different symbols correspond to the different values of the
   kernel halfwidth.}
   \label{fig8}
   \end{figure}

\begin{figure}
   \centering
   \includegraphics[width=8truecm]{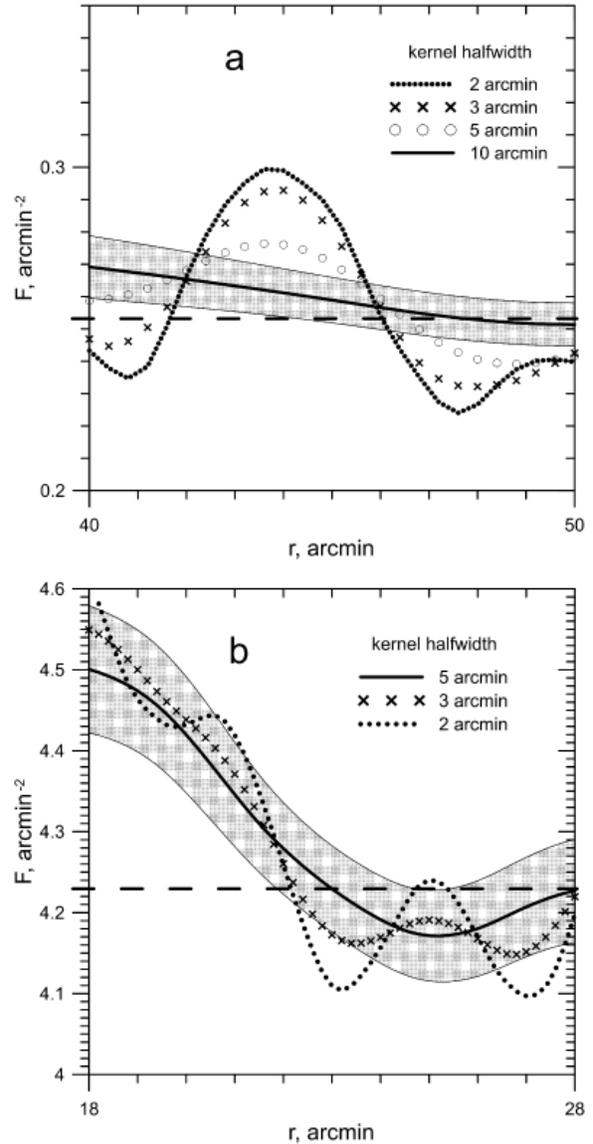}
   \caption{Surface density profiles of the clusters in the region around the cluster boundary,
   obtained with the different kernel halfwidth values. (a) NGC 2287,  $J_{\rm lim}=13$ mag;
   (b) NGC 6819, $J_{\rm lim}=16$ mag. Different symbols correspond to different values of the
   kernel halfwidth.  The horizontal dashed line shows the visual estimate of background density
   (see explanation below in Section 5). Grey bands show the $2\sigma$ confidence intervals
   for profiles with (a) $h=10$ and (b) $h=5$ arcmin.}
   \label{fig9}
   \end{figure}

\begin{figure}
   \centering
   \includegraphics[width=8truecm]{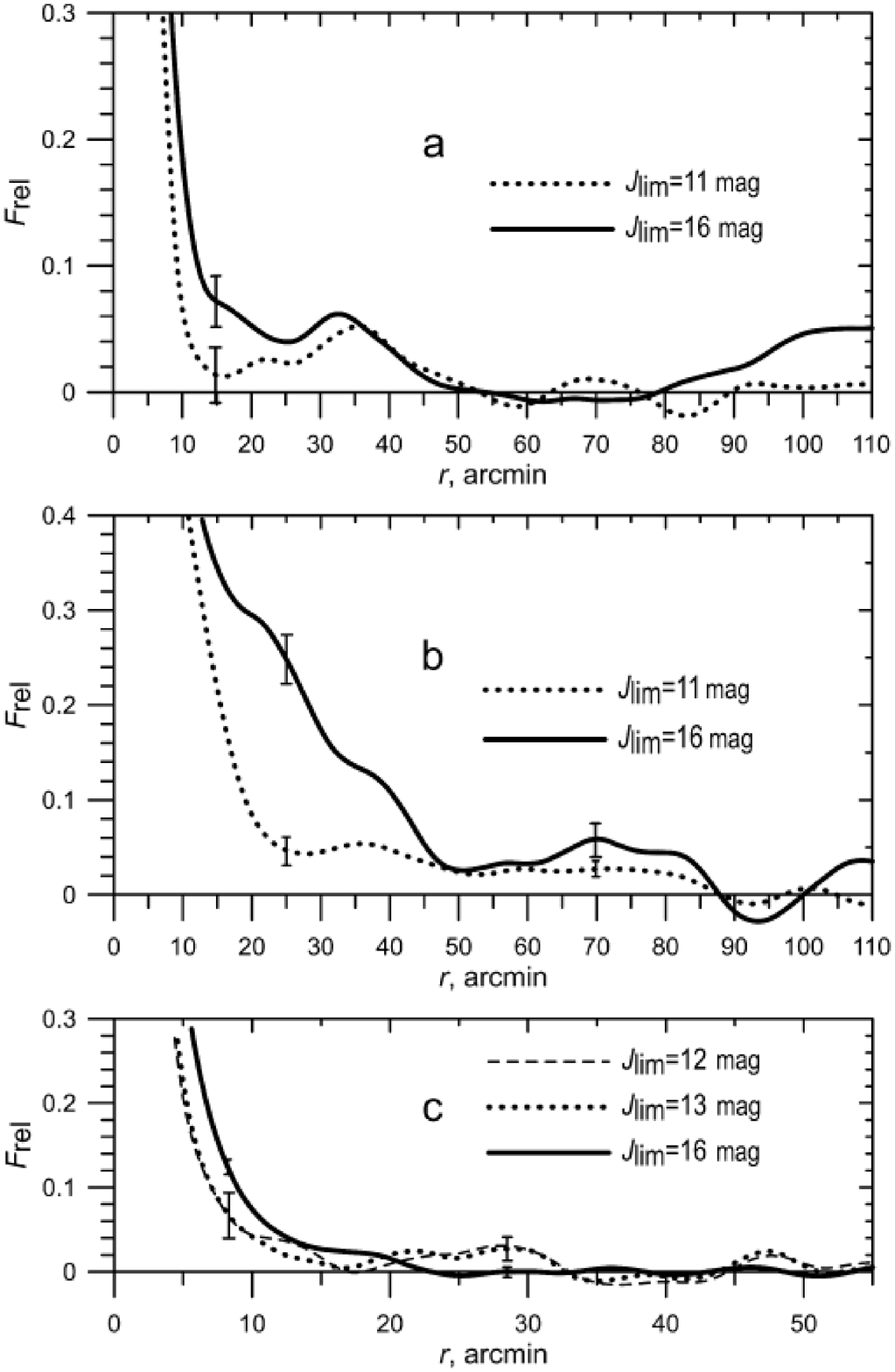}
   \caption{Comparing relative surface density profiles for different
   limiting magnitudes. a -- NGC 1502, b -- NGC 2516, c -- NGC 6819.
   Vertical bars show the width of the $2\sigma$ confidence interval.}
   \label{fig10}
   \end{figure}

Fig.6 shows the comparison of fixed and adaptive kernel estimates with the kernel halfwidth
$h=1$ pc (in the case of adaptive kernel estimator $h=1$ pc refers to the pilot estimate).
The adaptive kernel estimate was made with the restricted factors $\lambda$. The solid line
in this picture shows the adaptive kernel estimate of the spatial density in the corona of model
1 from \citet{DD} at the time point of about 150 Myrs
in the units of $\rm pc^{-3}$. The tidal radius of this model in the Galactic
gravitational field is of about 10 parsecs (see the formula for the tidal radius
below in Section 7). The dashed lines show the confidence interval
of $2\sigma$ width obtained by the smoothed bootstrap method (see \citet{MT}).
This method is based on the Monte--Carlo
simulation of multiple secondary samples. Secondary samples are created, which are
equal to the original one in size, and which are distributed in accordance with the
same density distribution as the original sample. Then the density estimate
for the every secondary sample is obtained, using the same kernel estimator. 20 secondary
samples were used in this work: it gave density dispersion values for every
$r_{\rm i}$ point. The fixed kernel estimate is shown by open circles.
It is clear, that adaptive kernel estimate with $h=1$ pc follows the mean trend defined by
the fixed kernel estimate with $h=1$ pc, and is relatively smooth.
The adaptive kernel estimate with $h=2$ pc has the same characteristics, but is smoother
in the central region. Adaptive estimates with $h=3, 4, \mbox{and}\; 5$ pc are biased
in the outer part of the corona model. Then the kernel halfwidths of 1 and 2 pc were
selected for estimation of spatial density of the open cluster corona model.

The evolution of the spatial density profile with time for the corona of cluster model 1 \citep{DD}
is shown in the sequences of frames `spatial density 1.flv' (the kernel halfwidth of 1 pc)
and `spatial density 2.flv' (the kernel halfwidth of 2 pc), which are accessible in the
online publication of this paper. Each frame is arranged as in Fig.6, but
without comparison with the fixed kernel estimate.
Each sequence contains 60 frames, the time interval is about 0.05 of
the violent relaxation time of this model \citep{DD}; that is, about 2.5 Myrs.
The last frame in the `spatial density 1.flv' is the same as Fig.6. It can be observed that
an imaginary upper envelope line for the density profile is stretched
to about three tidal radii of the model.
This confirms the results of \citet{DPS} on the formation of the
quasi-equilibrium density distribution in the cluster corona models. It means that the density
profile approaches with time to the upper envelope line which is just
the quasi-equilibrium density distribution. This temporal
equilibrium in the corona indicates a balance between the numbers of stars
entering the corona from inner regions of the cluster and escaping to the
corona periphery or beyond it \citep{DPS}.

\section{Surface density profiles for open clusters}

Surface density profiles for seven open clusters were obtained in this work
for different limiting magnitudes, $J_{\rm lim}$, with the data of 2MASS \citep{2MASS}. The sample
clusters are listed in Table 1. This table shows galactic coordinates of
clusters, their colour excesses, distance modules, distances and ages from \citet*{LGM},
with the last correction of the data (Loktin 2012, private communication).
With the exception of NGC 1960 all sample clusters were selected at large
galactic latitudes in order to have a more uniform and relatively low stellar
background density. Two clusters are young, two clusters are intermediate-aged
and three clusters are old. The cluster centre coordinates were taken from the
WEBDA database; their accuracy was found sufficient with the large kernel halfwidth
used in this work (usually 5 or 10 arcmin).

\begin{table*}
\normalsize
\bigskip
\begin{center}
\vspace{2 mm} Table 1. The sample clusters

\vspace{2 mm}
\begin{tabular}{|l|r|r|r|r|c|r|r|r|}
\hline
Cluster name   & $l$, deg & $b$, deg & E(B-V), mag    & Dist. mod., mag   & Distance, pc   &   Log age     & $h$, arcmin & $R_{\rm f}$, arcmin\\
\hline
NGC 1502       &  143.6   &    7.6   & 0.76$\pm$0.01  &  9.60$\pm$0.14    &  830$\pm$50    & 7.04$\pm$0.05 &    10       &       110          \\
NGC 1960 (M 36)&  174.5   &    1.0   & 0.23$\pm$0.04  & 10.59$\pm$0.10    & 1310$\pm$60    & 7.42$\pm$0.20 &     5       &        60          \\
NGC 2287 (M 41)&  231.1   &  -10.2   & 0.03$\pm$0.01  &  9.21$\pm$0.10    &  700$\pm$30    & 8.39$\pm$0.07 &    10       &       120          \\
NGC 2516       &  273.9   &  -15.9   & 0.10$\pm$0.01  &  8.10$\pm$0.11    &  420$\pm$20    & 8.10$\pm$0.04 &    10       &       110          \\
NGC 2682 (M 67)&  215.6   &   31.7   & 0.06$\pm$0.01  &  9.79$\pm$0.05    &  910$\pm$20    & 9.41$\pm$0.02 &     5       &       115          \\
NGC 6819       &   74.0   &    8.5   & 0.24$\pm$0.04  & 11.87$\pm$0.20    & 2360$\pm$200   & 9.17$\pm$0.07 &     5       &        55          \\
NGC 6939       &   95.9   &   12.3   & 0.33$\pm$0.03  & 10.45$\pm$0.36    & 1230$\pm$200   & 9.35$\pm$0.05 &    10       &       160          \\
\hline

\end{tabular}
\end{center}
\end{table*}

The case of the real open clusters is very different from the case of the open cluster N-body models.
The real clusters are observed at the rich stellar background, and the range of the estimates of the
surface density values in this case is much smaller, than the range of the estimates of the spatial (or surface)
density in the case of model. Due to this reason factors $\lambda$, that adjust kernel halfwidth
in the adaptive kernel algorithm, have small range also in the case of the real clusters. Factors $\lambda$
differ from the unity noticeably only in the region of the cluster core. As a result, the adaptive and
the fixed kernel estimates of the surface density differ only in the region of the cluster core
and coincide completely in the region of the cluster halo and corona. The present work is aimed
generally at the study of the outer regions of the open clusters, due to this reason the fixed
kernel algorithm is used in the present work for estimation of the surface density of the open clusters.

Let's examine how the result of surface density estimation depends on the kernel halfwidth $h$.
Fig.7 shows the radial surface density profiles for cluster NGC 2287 for $J_{\rm lim}=13$ mag,
obtained with the different kernel halfwidths.
It is seen that with the kernel halfwidth decrease the variation of profile increases.
Plots 7a, 7b, and 7c vary too greatly.
But it is difficult to estimate the degree of bias, because at the region of background
($r>60$ arcmin) all kernel halfwidths give the same estimate of background density value.
The comparison of the surface density estimates in the region, where the density gradient is changing
considerably (the outer part of the cluster core), is the best way for an estimation of the degree
of bias in that case. Fig.8 shows the surface density estimates for NGC 2287 obtained with the different kernel
halfwidths in the distance range $r\in[10,30]$ arcmin. It is seen that the curve with $h=10$ arcmin is smooth, and
follows well the mean trend defined by the curves computed with much smaller smoothing parameter.
The curves with the larger kernel halfwidths deviate from this trend appreciably.
Then the best value of the kernel halfwidth in this case is 10 arcmin, in accordance with
recommendations of \citet{MT}.

The same procedure was applied to all sample clusters for all values of the limiting
magnitude. One value of the kernel halfwidth was selected for every cluster, with the aim
of comparing the surface density estimates derived with different limiting magnitudes.
The last two columns of Table 1 show respectively the kernel halfwidth $h$ values accepted for the surface density radial profiles
construction of the sample clusters, and radii $R_{\rm f}$ of fields under consideration.
(Important note: in order to estimate the surface density by the kernel estimator
with the kernel halfwidth $h$ inside the circle of radius $R_{\rm f}$, the coordinates of
stars inside the circle with radius $R_{\rm f}+h$ are needed.)

Tables 2--8 contain data on the surface density profiles obtained in this work:
each table contains data for one cluster. All tables are accessible in the
online publication of this paper. All tables have the same organization;
an example of the first rows of Table 2 for NGC 1502 is given below. The first
column contains the distance from the cluster centre in arcmin. Columns 2--5 contain data
for limiting magnitude $J_{\rm lim}=11$ mag: column 2 is the kernel estimate of the surface density
radial profile with the kernel halfwidth listed in Table 1; column 3 is the
lower boundary of the confidence interval; column 4 is the upper boundary of the confidence
interval, column 5 is the surface density histogram with the bin width of 4 arcmin.
The histograms with the same bin width are tabulated for all clusters
(comparison of kernel estimates and histograms could be useful in some cases). Columns 6--9
contain the same data for limiting magnitude $J_{\rm lim}=12$ mag; columns 10--13
contain the same data for limiting magnitude $J_{\rm lim}=13$ mag; columns 14--17
contain the same data for limiting magnitude $J_{\rm lim}=14$ mag; columns 18--21
contain the same data for limiting magnitude $J_{\rm lim}=15$ mag; and columns 22--25
contain the same data for limiting magnitude $J_{\rm lim}=16$ mag. All surface density data
are in units of $\rm arcmin^{-2}$.

The surface density radial profiles for different limiting magnitudes are used
in the present work for estimation of the cluster masses, and for evaluation
of the segregation of the stars with the different masses (mass segregation).

The nominal completeness limit of 2MASS Point Source Catalogue is 15.8
mag \citep{2MASS}, but in the magnitude range
$J\in[15.8,16.0]$ mag this catalogue is 99\% complete for virtually all of the
sky \citep{2MASSes}. At the same time
the completeness limit is $\sim0.9$ mag fainter at high galactic latitude and
$\sim0.4$ mag brighter in the galactic plane \citep{2MASSes}. It means that the
completeness limit varies depending on the overall stellar density, and
the completeness in the last magnitude range ($J_{\rm lim}=16$ mag) can be
less than unity and different from one cluster to another.

\begin{table*}
\normalsize
\bigskip
\begin{center}
\vspace{2 mm} Table 2. Data on surface density radial profiles for NGC 1502.
The first ten columns and the first seven rows of the whole table, which is accessible in the
online publication of this paper.

\vspace{2 mm}
\begin{tabular}{|c|c|c|c|c|c|c|c|c|c|}
\hline
NGC 1502    &            &            &          &            &             &           &          &             &        \\
   r, arcmin&Jlim=11 mag &            &          &            &Jlim=12 mag  &           &          &             &        \\
        	&	   F     &  confidence& interval &  histogram &      F      & confidence& interval &  histogram  &        \\
\hline
       1    &      2     &      3     &     4    &      5     &      6      &     7     &     8    &     9       &   10   \\
\hline
     0.000  &   0.259154 &  0.223405  & 0.294902 &  0.497359  &   0.504841  & 0.446151  & 0.563532 &  0.875352   &   ...  \\
     0.200  &   0.258997 &  0.223273  & 0.294722 &  0.497359  &   0.504559  & 0.445912  & 0.563207 &  0.875352   &   ...  \\
     0.400  &   0.258527 &  0.222873  & 0.294180 &  0.497359  &   0.503711  & 0.445191  & 0.562231 &  0.875352   &   ...  \\
     0.600  &   0.257738 &  0.222203  & 0.293274 &  0.497359  &   0.502291  & 0.443980  & 0.560601 &  0.875352   &   ...  \\
     0.800  &   0.256633 &  0.221264  & 0.292003 &  0.497359  &   0.500293  & 0.442272  & 0.558314 &  0.875352   &   ...  \\
     1.000  &   0.255221 &  0.220064  & 0.290377 &  0.497359  &   0.497730  & 0.440076  & 0.555384 &  0.875352   &   ...  \\
     1.200  &   0.253507 &  0.218610  & 0.288405 &  0.497359  &   0.494615  & 0.437403  & 0.551827 &  0.875352   &   ...  \\
      ...   &            &            &          &            &             &           &          &             &        \\
\hline

\end{tabular}
\end{center}
\end{table*}

It may be seen from the results of \citet{MT}, that both kernel and
maximum likelihood methods overestimate the surface density in the region of the outer boundary
when large values of the smoothing parameter (the kernel halfwidth) are used for the restoration of the Plummer and
Michie--King distributions. In that case it is probable that the larger kernel halfwidth would
lead to larger cluster dimensions.

The real open clusters do not show the noticeable dependence of the cluster
radius on the kernel halfwidth, when the kernel halfwidths listed in Table 1 and
smaller ones are used. The possible explanation
is that open clusters are projected on a rich stellar background as opposed to the
\citet{MT} models, where stellar background is not taken
into account. This is illustrated in Fig.9. Fig.9a shows surface
density profiles in the region around the cluster boundary for cluster NGC 2287
for $J_{\rm lim}=13$ mag, for kernel halfwidth values
of 2, 3, 5 and 10 arcmin. Fig.9b shows surface
density profiles in the region around the cluster boundary for cluster NGC 6819
for $J_{\rm lim}=16$ mag, for kernel halfwidth values
of 2, 3, and 5 arcmin.
The cluster boundary (the value of the cluster radius) is determined by
the intersection of the cluster surface density profile, obtained with the
kernel halfwidth listed in Table 1 and marked in Fig.9 by the thick solid lines, with the line of background density
(the dashed line, see explanation below in Section 5). It is clearly noted, that intersection points
of the other surface density profiles (obtained with the smaller kernel halfwidth values)
with the background density line (near 46--47 arcmin
in Fig.9a, and near 22--23 arcmin in Fig.9b) are inside the bands of the confidence
interval for profiles with the kernel halfwidth values from Table 1 (the larger ones).

Density profiles obtained with different limiting magnitudes were
compared in the present work in order to find the signs of mass segregation
in the sample clusters. As the surface density values differ
greatly for different limiting magnitudes, relative
densities were used, determined by the following formula, where $F_{\rm b}^{\rm vis}$ is the
visual estimate of the surface density of the stellar background
(see explanation in Section 5), and $F(0)$ is the surface density in
the cluster centre:

\begin{equation}
\label{reldens}
F_{\rm rel}(r_{\rm i})=\frac{F(r_{\rm i})-F_{\rm b}^{\rm vis}}{F(0)-F_{\rm b}^{\rm vis}} \; \mbox{.}
\end{equation}

The comparison of the relative density profiles for clusters NGC 1502,
NGC 2516 and NGC 6819
is shown in Fig.10: Fig.10a is for NGC 1502; Fig.10b is for NGC 2516;
and Fig.10c is for NGC 6819.
Two types of differences can be marked. The first one is presented
in all three clusters: the outer part of the cluster
core (or `intermediate zone') is relatively more populous
in faint stars. The second type is seen in the case of NGC 2516,
where the cluster halo is also more populous in faint
stars. All sample clusters show differences of one type or the other.
In all cases the relative population of faint stars
in the outer cluster regions exceeds the relative population of
brighter stars, apart from NGC 6819, where the opposite
picture can be seen (Fig.10c).

The Kolmogorov-Smirnov test (KS-test) was performed in order to statistically compare
the relative density profiles in Fig.10 \citep{Num_Rec}.
For the profiles from Fig.10a KS-test gives the p-value of $3.8\cdot10^{-3}$,
and for the profiles from Fig.10b -- $3.4\cdot10^{-10}$. That is, these profiles
are statistically different. For the profiles from Fig.10c, KS-test gives
the following results. The profiles with $J_{\rm lim}=12$ mag and
$J_{\rm lim}=13$ mag are not statistically different (the corresponding
p-value is 0.9999). The profiles with $J_{\rm lim}=12$ mag and
$J_{\rm lim}=16$ mag are statistically different (the corresponding
p-value is $4.1\cdot10^{-7}$). The profiles with $J_{\rm lim}=13$ mag and
$J_{\rm lim}=16$ mag are also statistically different (the corresponding
p-value is $1.3\cdot10^{-6}$).

The mass of sample cluster stars for different magnitudes can be estimated.
Transition to absolute magnitudes $M_J$ was made with the data on
cluster distances and colour excesses $E(B-V)$ from \citet{LGM} catalogue
and with the use of the formulae:

\begin{equation}
\label{excess}
E(J-H)=0.37E(B-V) \; \mbox{, and}
\end{equation}

\begin{equation}
\label{extinction}
A_{\rm J}=2.43E(J-H) \; \mbox{,}
\end{equation}

\noindent where $E(J-H)$ is the colour excess in $(J-H)$ colour index,
and $A_{\rm J}$ is the total extinction in $J$ colour.
Formula (\ref{excess}) was taken from \citet{BB};
formula (\ref{extinction}) from \citet{LS}.
Then, the masses of stars were estimated by their absolute magnitudes $M_J$
with isochrone tables downloaded from http://stev.oapd.inaf.it/cmd
\citep{BMGSDRN} with $Z_\odot=0.019$. The isochrone of $\lg t=7.0$ was
used for clusters NGC 1502 and NGC 1960; the isochrone of $\lg t=8.3$ was used
for clusters NGC 2287, NGC 2516; and the isochrone of $\lg t=9.3$ was used
for clusters NGC 2682, NGC 6819, and NGC 6939.
One isochrone is used for the young clusters, one isochrone for the
intermediate-aged clusters, and one isochrone for the old clusters.
The reason is that only mass--luminosity relation is important in the
present work, and this relation changes only negligibly for isochrone
with close age values. It is important, that this method
does not require the matching of the isochrone to the cluster
colour-magnitude diagram (CMD).

The data on stellar masses corresponding to stellar magnitudes
in the sample clusters are listed in Table 9. $J_{\rm up}$ in this
table denotes the magnitude of the upper end of the cluster sequence
in the CMD. In order to find this value, the CMDs ($J,(J-H)$) for sample clusters
were plotted by the data of 2MASS in the regions of 10 arcmin
around the cluster centre. The uncertainties in this table are due to
uncertainties in the cluster distance modules, and in the colour
excesses for the clusters (see Table 1). Where the uncertainty interval
was determined as asymmetric, the larger value was listed.

\begin{table*}
\normalsize
\bigskip
\begin{center}
\vspace{2 mm} Table 9. The stellar masses at the boundaries
of magnitude intervals in the sample clusters ($M_\odot$).
$J_{\rm up}$ is the magnitude of the upper end of the cluster sequence
in the CMD (see explanation in the text).

\vspace{2 mm}
\begin{tabular}{|l|r|r|r|r|r|r|r|}
\hline
Cluster name   &   $J_{\rm up}$  & $J=11$ mag    & $J=12$ mag    &     $J=13$ mag&   $J=14$ mag  &     $J=15$ mag&    $J=16$ mag  \\
\hline
NGC 1502       & 17.31$\pm$0.29  & 3.35$\pm$0.23 & 1.91$\pm$0.33 & 1.43$\pm$0.03 & 1.15$\pm$0.05 & 0.74$\pm$0.06 & 0.40$\pm$0.05  \\
NGC 1960 (M 36)& 11.15$\pm$0.38  & 4.29$\pm$0.22 & 2.72$\pm$0.14 & 1.53$\pm$0.02 & 1.32$\pm$0.03 & 0.97$\pm$0.05 & 0.59$\pm$0.05  \\
NGC 2287 (M 41)&  4.09$\pm$0.00  & 1.95$\pm$0.08 & 1.37$\pm$0.04 & 1.07$\pm$0.03 & 0.83$\pm$0.03 & 0.65$\pm$0.02 & 0.49$\pm$0.02  \\
NGC 2516       &  3.87$\pm$0.00  & 1.35$\pm$0.04 & 1.06$\pm$0.03 & 0.82$\pm$0.03 & 0.64$\pm$0.02 & 0.49$\pm$0.02 & 0.33$\pm$0.02  \\
NGC 2682 (M 67)&  1.72$\pm$0.00  & 1.67$\pm$0.01 & 1.43$\pm$0.01 & 1.18$\pm$0.02 & 0.94$\pm$0.01 & 0.74$\pm$0.01 & 0.59$\pm$0.01  \\
NGC 6819       &  1.72$\pm$0.00  & 1.71$\pm$0.00 & 1.71$\pm$0.00 & 1.70$\pm$0.02 & 1.50$\pm$0.06 & 1.24$\pm$0.05 & 1.00$\pm$0.05  \\
NGC 6939       &  1.72$\pm$0.00  & 1.71$\pm$0.01 & 1.65$\pm$0.08 & 1.41$\pm$0.10 & 1.15$\pm$0.09 & 0.92$\pm$0.08 & 0.73$\pm$0.06  \\
\hline

\end{tabular}
\end{center}
\end{table*}

The differences in the relative density profiles with the different limiting
magnitudes are present in all sample clusters. It is seen from Table 9
that, at least in the young and intermediate-age clusters, there is a
large mass spectrum: then we can explain the differences in the profiles
there as the consequence of a mass segregation process. In the case of
NGC 6819 the outer part of the cluster core is more populated with
faint stars, but the cluster halo is more populous with the
brighter stars. However, the difference in the mass between cluster
stars in that case is minimal, and this fact has yet to be interpreted.

\begin{figure*}
   \centering
   \includegraphics[width=18truecm]{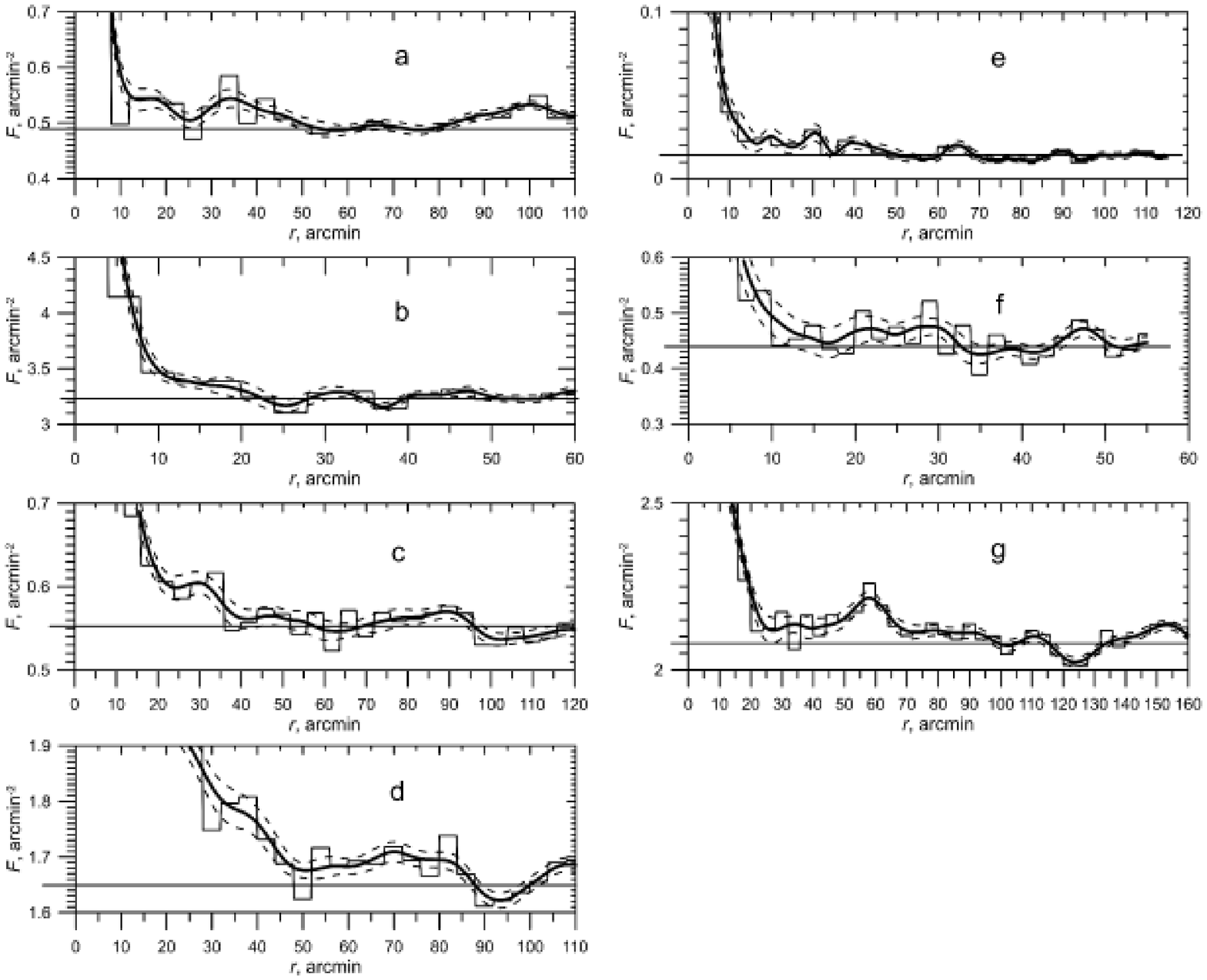}
   \caption{Structural irregularities in the surface density profiles
   of open clusters. a -- NGC 1502 $J_{\rm lim}=14$ mag, b -- NGC 1960 $J_{\rm lim}=16$ mag, c -- NGC 2287 $J_{\rm lim}=14$ mag,
   d -- NGC 2516 $J_{\rm lim}=16$ mag, e -- NGC 2682 $J_{\rm lim}=11$ mag, f -- NGC 6819 $J_{\rm lim}=13$ mag,
   g -- NGC 6939 $J_{\rm lim}=16$ mag. The solid polygonal lines show the histograms with the bin size of 4 arcmin.
   The thick solid lines show the surface density estimate, the dashed lines show confidence interval of $2\sigma$ width.
   The solid straight lines show the values of stellar density of background (see explanation in Section 5). }
   \label{fig11}
   \end{figure*}

The sample clusters show the presence of structural irregularities
in their density profiles, such as secondary maxima or `footsteps'
('footstep' is the same as 'plateau').
The only exception is NGC 1960. The examples are shown in Fig.11.
The typical `footstep' is seen in NGC 2287 near $r=30$ arcmin, the typical secondary
maximum is seen in NGC 6939 near $r=60$ arcmin.
Such structures can indicate the cluster non-stationarity in the
regular field, or stabilizing ejections of the cluster stars into the
galactic field: see \citet{D82,D05,D11}. The non-stationary processes cause
the corona not being radially symmetric, and this, in turn, leads again to
the structural irregularities in the radial density profiles.

\section{Sizes of open clusters}
The sizes of open clusters were estimated in the present work in
two ways. The first one was by a visual estimate, and it was not an objective method.

In the first step,
the mean background surface density line was inferred by analysing
the outer part of the field under consideration for every cluster,
and for every limiting magnitude range. An approximately flat area
in the outer part of the density profile was searched, and the
background density line was drawn taking into account an approximate
equality of the square of areas between this line and density profile
above and below this line. In the second step the cluster
radius was estimated as the abscissa of the point of intersection
of the density profile and the background density line. An error of
this estimate was evaluated as the distance from the intersection
point of the confidence interval line with the background
density line to the cluster radius point (in many cases the confidence
interval intersects the background density line only at one side
of the cluster radius point). An error of the background density
estimate was evaluated as half of the confidence interval width
at the cluster radius point.

These background density lines are shown in Fig.9 and Fig.11.
The visual estimates of the cluster radius $R_{\rm c}$ and the surface density
of stellar background $F_{\rm b}^{\rm vis}$, and their uncertainties for every cluster and
every limiting magnitude interval are listed in
Table 11. The intervals of the cluster radius estimates for every cluster
are listed in the second column of Table 10.

The second way is the approximation of the cluster surface density
profile by the King surface density distribution (King 1962), and by
the combination of the King distribution and the cluster corona component
(see the description and the discussion in Section 6).
It is important that the visual estimates of the
mean background surface density and the estimates of the background density
via approximation with the combined function are very close
(see Table 11).

Table 10 shows the comparison of visual estimates of open cluster
radii both with the data of other authors and with the results of
cluster radii estimation by the DMP method when the function $N(r)$
(number of stars in the circle with radius $r$) is used, and the
cluster field is compared with several fields of neighbouring
background fields (see above in Introduction). All data in Table 10 are in arcmin.

\begin{table*}
\normalsize
\bigskip
\begin{center}
\vspace{2 mm} Table 10. Comparison of cluster radii estimates
with the data of other authors and with the results of
cluster radii estimation by the DMP method (arcmin)

\vspace{2 mm}
\begin{tabular}{|l|r|r|r|r|r|r|}
\hline
Cluster name   & Cluster radius  & Kharchenko   & Data of & Ref. & Danilov \& Seleznev & Radii estimates   \\
               & estimate by     & et al. (2005)& other   &      & (1994) DMP method   & by DMP method     \\
               & density profile & catalog      & authors &      & with plates in B    & with 2MASS    \\
\hline
NGC 1502       &  52-55  (110)   &   12.6       &   5     &  1   & $24.8\pm2.5$ (31.08) & 37 (45)   \\
NGC 1960 (M 36)&  10-23   (60)   &   16.2       &   22.9  &  2   & $20.1\pm0.6$ (31.08) &           \\
NGC 2287 (M 41)&  37-57  (120)   &   30         &   30    &  3   &                      & 46-50 (60)\\
NGC 2516       &  88-92  (110)   &   42         &   90    &  3   &                      & 87 (95)   \\
NGC 2682 (M 67)&  43-57  (115)   &   18.6       &   60    &  4,5 &                      &           \\
NGC 6819       &  16-33   (55)   &              &   13    &  6   & $24.8\pm2.6$ (31.08) & 10-22 (40)\\
NGC 6939       &  42-105 (160)   &              &   85    &  7   & $15.5\pm1.2$  (22.2)  & 21-26 (30)\\
\hline
\multicolumn{7}{l}{\footnotesize  (1) \citet{APKB}, (2) \citet{SA}, (3) \citet*{BLG}, (4) \citet{DS}, } \\
\multicolumn{7}{l}{\footnotesize  (5) \citet{Esp}, (6) \citet{YSDSKK}, (7) \citet{AK} } \\

\end{tabular}
\end{center}
\end{table*}

The second column of Table 10
contains the visual estimates of cluster radius by the surface density profile
obtained as described above.
The interval shows the scatter of the estimates for the different limiting magnitudes.
The number in brackets is the radius of the field used for the density profile
construction. The third column shows the cluster radius from the catalogue
of \citet{K2004}. The fourth column shows the data on the sample
clusters from the literature, and the fifth column contains the references
on the sources of these data.
The sixth column contains the cluster radius estimates from \citet{DS94}.
These estimates were obtained by the DMP method with star counts
on photographic plates in B colour band. The number in brackets shows the radius
of the cluster field used for the star counts. The seventh column shows the cluster
radius estimates obtained by the DMP method with the star counts on the data
of 2MASS. The interval shows the scatter of estimates for different
limiting magnitudes, and the number in a brackets shows the radius
of the cluster field used for the star counts.

The radius estimates by the surface density profile
in the case of NGC 1502, NGC 6819, and NGC 6939 are
larger than estimates by star counts with the DMP method.
This can be explained by a smaller size of the cluster field used for the DMP star counts.
In the case of NGC 1960, NGC 2287, and NGC 2516 the size of the field
used for the star counts with the DMP method is larger than the cluster size,
and a satisfactory matching by different methods was obtained.

It may be seen from Table 10 that, in the case of NGC 1502, NGC 2287, and
NGC 6819, we have in the literature underestimated values of the cluster radius.

\citet{AK} studied the structure of NGC 6939
with the proper--motion--selected cluster members.
They found that this cluster has an extensive corona
with the radius of about 85 arcmin. In the present work, the surface density
profile for NGC 6939 was derived to a distance of 160 arcmin from
the cluster centre, and the cluster radius estimate
larger than in \citet{AK} was obtained: see Fig.11g. In this manner the result
of \citet{AK} concerning an extensive corona of NGC 6939 can be confirmed.
The cluster radius estimate
comparable with the result of proper motion cluster membership analysis
was obtained in the case of NGC 2682 \citep{Esp}.
\citet{K2004} used proper motions data for selecting possible
cluster members, but these authors obtained smaller cluster radii than in the
present work. This is possibly due to the smaller limiting magnitude in
their study, and possibly due to using the \citet{King} distribution
for the cluster structure approximation (see discussion in Section 6).

\citet{Nilakshi} performed star counts in the fields of 38 open clusters.
They obtained the outer radius of NGC 1960 to be 15.3 arcminutes and the
outer radius of NGC 6939 to be 12.7 arcminutes (these values of angular radii
were calculated with their data on linear radii and distances). These radii
are smaller than the ones obtained in the present paper. In the case of NGC 6939
\citet{Nilakshi} couldn't see the cluster boundary near 100 arcminutes,
because they were limited by the field with the radius of 30 arcminutes.
Their result must be compared rather with \citet{DS94} value, see 6th column
of Table 10. In the case of NGC 1960 the reason of underestimation of the
radius by \citet{Nilakshi} is possibly due to the lower sensitivity of star counts
in the rings in comparison with the kernel estimator method. It is worthy
to note that the procedure of the outer boundary determination was not
described by \citet{Nilakshi} in details, and the density profiles (see
Fig.1 in \citet{Nilakshi}) allow ambiguous radii estimation.

\section{Approximation of open cluster surface density profiles}

The \citet{King} function is used very often for approximation of
the surface density or the surface brightness profiles of star clusters:

\begin{equation}
\label{King}
F(r)=\left\{
	    \begin{array}{ll}
	     k\Biggl[\frac{\displaystyle 1}{\displaystyle \sqrt{1+\left(\frac{r}{r_{\rm c}}\right)^2}}-\frac{\displaystyle 1}{\displaystyle \sqrt{1+\left(\frac{r_{\rm t}}{r_{\rm c}}\right)^2}}\Biggr]^2 & \: r<r_{\rm t} \; \mbox{,}\\
         &    \\
	    0 & \: r \ge r_{\rm t} \; \mbox{.}\\
	    \end{array}
	   \right.
\end{equation}

This function was proposed by King for globular clusters but was
also widely used for open clusters. In order to take into account
stellar background, this formula is supplemented by
stellar background density $F_{\rm b}$ as a constant addition.

\citet{DP} found that the approximation of stellar distribution
in open star clusters by the \citet{King} function tends to underestimate
the number of stars in the cluster compared to the results of star counts.
The reason was that the \citet{King} function underestimates density values
in the region of the cluster corona.
\citet{DP} proposed an addition to the King formula.
This addition represents the cluster corona as a uniform sphere.
The addition into surface density reads:

\begin{equation}
\label{unisphere}
\delta F(r)=2\cdot R_2\cdot\delta f\cdot \sqrt{1-\left(\frac{r}{R_2}\right)^2} \; \mbox{.}
\end{equation}

\noindent where $R_2$ is the radius of the cluster corona, and $\delta f$ is
the spatial density of the cluster corona. This addition should be applied at
all radii $r<R_2$.

An approximation of the surface density profiles of the sample clusters
was performed in the present work both by the \citet{King} function alone (Eq.(\ref{King}),
referred hereafter as `King model') and by the combined function (a combination of the King distribution
for the cluster core Eq.(\ref{King}) and of the uniform sphere Eq.(\ref{unisphere}) for the cluster
corona, referred hereafter as `combined model').

The results of the approximation are listed in Table 11, which is accessible in the
online publication of this paper. The columns of the table can be divided into three groups.
The first group contains visual estimates of the cluster parameters,
the second group contains the parameters of the combined model, and
the third group contains the parameters of the King model.

The columns of the first group are: (1) the cluster name;
(2) the limiting magnitude in \emph{J} band; (3) visual estimate of the cluster radii $R_{\rm c}$
in arcmin; (4) its uncertainty;
(5) visual estimate of the surface density of
the stellar background $F_{\rm b}^{\rm vis}$ in $\rm arcmin^{-2}$; (6) its uncertainty;
(7) the estimate of the cluster star number $N$; (8) its uncertainty. The estimate
of the cluster star number was obtained
through the numerical integration of the cluster surface density profile; the uncertainty
of this estimate was obtained by integration of the upper and lower
confidence interval curves, taking into account the uncertainty
in the background density.

The parameters of the combined model were
obtained by using the non-linear least-square approximation algorithm
by \citet{Marq}. The parameters of Eq.(\ref{King}) in the case of the combined model are supplied by the upper
index `comb', and in the case of the King model -- by the upper index `King'.
The columns of the second group are: (9) $k^{\rm comb}$ in $\rm arcmin^{-2}$;
(10) its uncertainty; (11) $r_{\rm c}^{\rm comb}$ in arcmin; (12) its uncertainty;
(13) $r_{\rm t}^{\rm comb}$ in arcmin; (14) its uncertainty;
(15) the surface density of background $F_{\rm b}^{\rm comb}$ in $\rm arcmin^{-2}$;
(16) its uncertainty; (17) $R_2$ in arcmin; (18) its uncertainty;
(19) $\delta f$ in the units of $10^{-3}\rm arcmin^{-3}$ (this
value denotes the number of stars in a cube with the side
measured by one arcmin at the cluster distance); (20) its uncertainty. In the combined
model, $r_{\rm t}^{\rm comb}$ can be considered as the cluster core radius, $r_{\rm c}^{\rm comb}$
has the meaning of the scale parameter for the cluster core,
and $R_2$ is the cluster corona radius. From this perspective,
situations when $r_{\rm c}^{\rm comb}>r_{\rm t}^{\rm comb}$ (see Table 11) are possible.
The interpretation of such cases is in the different types of the surface density
profiles, namely, in the differences in the transition region
between the cluster core and the cluster corona (or the halo). The cluster
can have the so-called intermediate zone between the core and the corona
\citep{K_str,DS94}. The existence of the intermediate zone is normal
in rich clusters \citep{K_str}, and the sample clusters are rather rich.
When the intermediate zone exists, the relation of $r_{\rm c}^{\rm comb}$
and $r_{\rm t}^{\rm comb}$ is usual. But when the transition between the core and the corona
is sharp, the scale parameter for the cluster core is larger than the radius of the core.
Such cases occur only in the less populated clusters of the sample, NGC 1502 and NGC 2287.

The following columns of the second group are: (21) the chi-square parameter
describing the approximation quality \citep{Marq,Num_Rec};
(22) the cluster star number $N_{\rm mod}$
for the combined model obtained by the analytic expression for the integral of Eq.(\ref{surfdens})
over the surface density of the combined model $[F(r)+ \delta F(r)]$ (see Eq.(\ref{King}), Eq.(\ref{unisphere}));
(23) the star number of the cluster corona $N_1$; and (24) the star number of the cluster core $N_2$.
The number of the cluster corona stars $N_1$ was obtained
by the analytic expression for integral Eq.(\ref{surfdens}) over the surface density
of cluster corona Eq.(\ref{unisphere}). The number of the cluster
core stars was obtained as $N_2=N_{mod}-N_1$.

The third group of the columns of Table 11 lists parameters of the King model
obtained for the sample clusters by the same algorithm \citep{Marq}:
(25) $k^{\rm King}$ in $\rm arcmin^{-2}$; (26) its uncertainty;
(27) $r_{\rm c}^{\rm King}$ in arcmin; (28) its uncertainty; (29) $r_{\rm t}^{\rm King}$ in arcmin;
(30) its uncertainty; (31) $F_{\rm b}^{\rm King}$ in $\rm arcmin^{-2}$;
(32) its uncertainty; (33) the chi-square parameter;
(34) the cluster star number $N_{\rm King}$
for the King model obtained by the analytic expression for integral Eq.(\ref{surfdens})
over the surface density of the King model, Eq.(\ref{King}).

The results of the approximation by two models are now compared.
The parameter $R_2$ in the combined model
correlates closely with the visual estimate of the cluster radii $R_{\rm c}$.
In contrast, parameter $r_{\rm t}$ in the King model does not correlate highly with $R_{\rm c}$.
It is shown in Fig.12.

\begin{figure}
   \centering
   \includegraphics[width=8truecm]{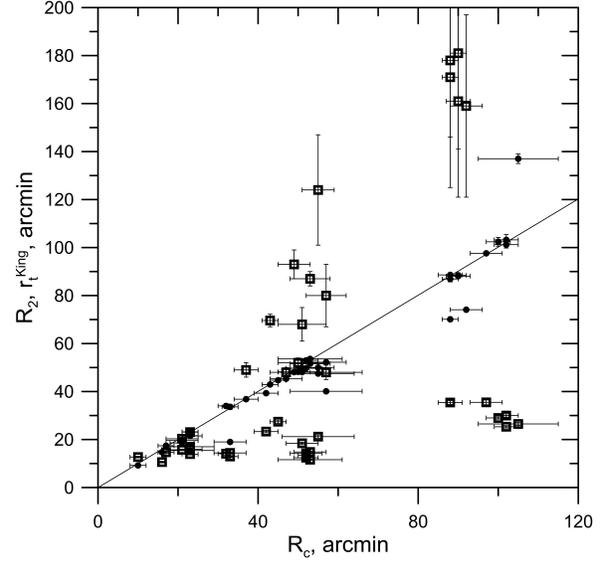}
   \caption{Comparison of the values $R_2$ and $r_{\rm t}^{\rm King}$ against the $R_{\rm c}$ values.
   The filled circles are $R_2$ values, and open squares are $r_{\rm t}^{\rm King}$ values.
   The straight line shows equal values, for convenience. }
   \label{fig12}
   \end{figure}

Stellar background density $F_{\rm b}^{\rm King}$, obtained in the limits of the King
model, is usually larger than $F_{\rm b}^{\rm comb}$ obtained in the limits of the combined
model (the latter one is usually very close to the visual estimate of this value).
It is clear, when corresponding columns of Table 11 are compared.

One could compare the relative differences of the surface densities of background.
The relative difference $(F_{\rm b}^{\rm vis}-F_{\rm b}^{\rm comb})/F_{\rm b}^{\rm vis}$
is generally smaller than 1 percent and not more than 4 percent.
The relative difference $(F_{\rm b}^{\rm comb}-F_{\rm b}^{\rm King})/F_{\rm b}^{\rm comb}$
is generally several times larger in the absolute magnitude, and usually negative.
%

The reason is that the King model does not have an extended corona,
and the cluster corona (that is seen clearly in the above Fig. 10 and Fig.11) is
perceived by the approximation algorithm as part of the stellar background.
Fig.13 shows the surface density profile for NGC 1502 ($J_{\rm lim}=16$ mag),
and the fits of this profile both by the King model and by the combined model.
It is visible, that the fit by the King model gives the values of the surface density
at the distances from the cluster centre between 50 and 80 arcmin
(in the background region) larger than the profile values, in contrast
to the fit by the combined model. As a result, integration of
the density profile of the cluster King model gives a number of stars $N_{\rm King}$
much smaller than $N$ or $N_{\rm mod}$: usually $N_{\rm King}$ is close to
the cluster core star number $N_2$ in the combined model.
In contrast, values of $N$ and $N_{\rm mod}$ are well correlated.
This fact is illustrated in Fig.14, where the cluster star numbers
in the combined model and in the King model are compared against
the cluster star number from the visual estimate of parameters.

\begin{figure}
   \centering
   \includegraphics[width=8truecm]{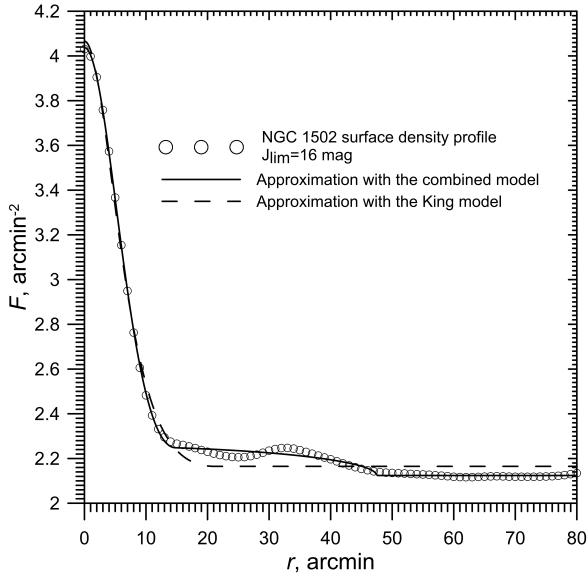}
   \caption{Approximation of surface density profile of NGC 1502 with $J_{\rm lim}=16$
   by the combined function and the King function.}
   \label{fig13}
   \end{figure}

\begin{figure}
   \centering
   \includegraphics[width=8truecm]{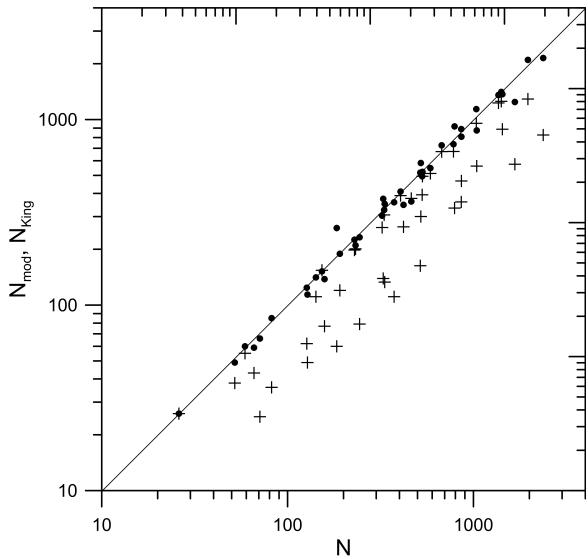}
   \caption{Comparison of the values $N_{\rm mod}$ (the cluster star number in the combined model)
   and $N_{\rm King}$ (the cluster star number in the King model) against the
   values of $N$ (the cluster star number from the visual estimate of parameters),
   shown for different limiting magnitudes for each sample cluster.
   The filled circles are $N_{\rm mod}$ values, and crosses are $N_{\rm King}$ values. }
   \label{fig14}
   \end{figure}

\begin{figure}
   \centering
   \includegraphics[width=8truecm]{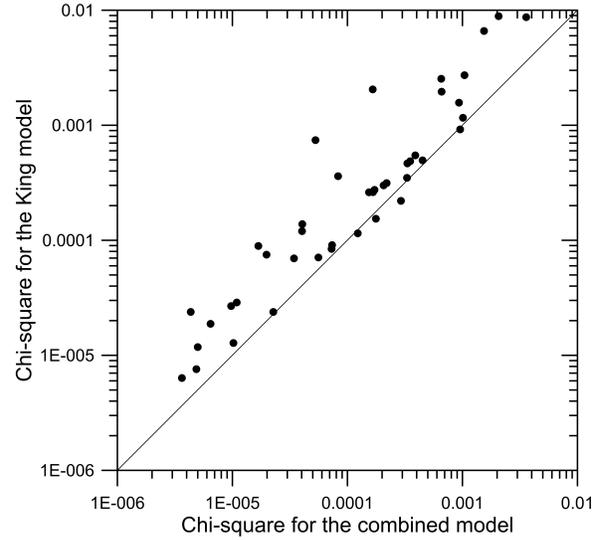}
   \caption{Comparison of the chi-square parameters for the King model approximation
   against the chi-square parameters for the combined model approximation. }
   \label{fig15}
   \end{figure}

Hence, it follows that the King model does not reproduce
surface density profiles of the sample clusters very well.
This point is supported by the comparison of the chi-square parameters,
describing the quality of approximation \citep{Marq,Num_Rec}.
Fig.15 shows the chi-square parameters for the King model approximation
against the chi-square parameters for the combined model approximation,
the latter ones are systematically less.
The cluster cores are reproduced by the King function accurately, but the cluster
coronae are not. Taking into account that the cluster coronae often have
structural irregularities (see Fig.11), it is difficult to
reproduce their density profiles by any analytic expression.
From this point of view modelling of the cluster corona by a
uniform sphere can be reasonable, and gives acceptable results.

\section{Cluster mass and tidal radii estimates}
Having data on the cluster star numbers and on the stellar masses
at the boundaries of magnitude intervals, it is possible to estimate
the cluster masses.
The following algorithm was used. First, the numbers of cluster
stars for magnitude intervals of 1 mag width were calculated (and their uncertainties).
Then these numbers were multiplied by the mean stellar masses obtained
from the data of Table 9, for every magnitude interval.
The mass of the cluster stars from the
upper magnitude interval was estimated with the assumption
of the Kroupa mass spectrum \citep{Kroupa} in this interval (see below in this Section). Finally, the cluster
mass estimates were obtained as the sum of the masses for all
magnitude intervals. The obtained cluster masses are the lower estimates,
because the unknown low-mass end
of stellar mass distribution, unresolved binaries and probable
remnants of massive stars, are not taken into account. These lower estimates of the sample
cluster masses are listed in the second column of Table 12. In the case of NGC 2287,
the estimate of its mass was carried out only up to $J_{\rm lim}=15$ mag,
because in the case of NGC 2287 the cluster star number with $J_{\rm lim}=16$ mag
is smaller, than the cluster star number with $J_{\rm lim}=15$ mag (see Table 11).
This fact can be explained by the large-scale fluctuations of the stellar background
density. It could result in the wrong (higher) estimate of the surface density
of the stellar background, and, as a consequence, in the wrong (lower) estimate
of the cluster star number in the case of $J_{\rm lim}=16$ mag.

\begin{table*}
\normalsize
\bigskip
\begin{center}
\vspace{2 mm} Table 12. Lower estimates of the sample cluster masses and tidal radii

\vspace{2 mm}
\begin{tabular}{|l|c|c|c|c|}
\hline
Cluster name   & Lower estimate  & Lower estimate   &$R_{\rm c\:max}$,& $R_{\rm 2\:max}$, \\
               & of cluster mass & of tidal radius  &  pc             &  pc               \\
               & $M$, $M_\odot$  & $R_{\rm t}$, pc  &                 &                   \\
\hline
NGC 1502       &  1300$\pm$140   &   14.1$\pm$1.2   &  13.3$\pm$2.2   &  12.9$\pm$0.2     \\
NGC 1960 (M 36)&  860$\pm$100    &   12.3$\pm$1.0   &  8.8$\pm$1.1    &  8.8$\pm$0.2      \\
NGC 2287 (M 41)&  880$\pm$150    &   12.6$\pm$1.2   &  11.6$\pm$1.8   &  9.8$\pm$0.1      \\
NGC 2516       &  1820$\pm$200   &   15.4$\pm$1.3   &   11.2$\pm$0.5  &  10.8$\pm$0.04    \\
NGC 2682 (M 67)&  1400$\pm$110   &   15.1$\pm$1.2   &  15.1$\pm$1.3   &  13.8$\pm$0.2     \\
NGC 6819       &  1890$\pm$140   &   16.7$\pm$1.3   &  22.7$\pm$2.7   &  23.3$\pm$0.7     \\
NGC 6939       &  2610$\pm$420   &   18.3$\pm$1.7   &  37.6$\pm$3.6   &  49.0$\pm$0.7     \\
\hline

\end{tabular}
\end{center}
\end{table*}

The total cluster mass, that was not covered by the method
adopted here, can be estimated. NGC 1502 is taken as the only example.
The following assumptions and approaches were used.

1. The mass interval for stars included into star counts is [0.4; 17.3]
solar masses. These values are taken from Table 9.
The mass interval for low-mass (unseen) stars is [0.08; 0.4] solar masses.
The initial mass interval of the massive stars, finished their evolution already,
is [17.3; 60.] solar masses.

2. Kroupa initial mass spectrum \citep{Kroupa} is adopted for these mass intervals:
$$
\phi(m) \sim \left\{
	    \begin{array}{ll}
	    m^{-1.3\pm0.5} & \mbox{with}\quad m\in[0.08;0.5] \; \mbox{,}\\
	    m^{-2.3\pm0.3} & \mbox{with}\quad  m>0.5  \; \mbox{.}\\
	    \end{array}
	   \right.
$$

3. The number of the stars in the mass interval of $[m_1;m_2]$ is
$$
N=\int\limits_{m_1}^{m_2}\phi(m)dm \; \mbox{,}
$$
the mass of the stars in the same mass interval is
$$
M=\int\limits_{m_1}^{m_2}m\phi(m)dm \; \mbox{.}
$$

4. The normalization constant of the Kroupa initial mass spectrum
is determined, because the number of the cluster
stars in the mass range of [0.4,17.3] (taken from Table 9) is 860 (Table 11).

5. The open cluster NGC 1502 is young (see Table 1),
and the fraction of low-mass stars lost by the cluster due to relaxation is
negligible, see, for example, \citet{EBJN}. For the intermediate-aged and old clusters
the star escapes should be considered, but the procedure of the total mass
evaluation was applied to NGC 1502 only, as the example.

6. The stars with the initial masses within the range of [17.3; 60.] solar masses become the neutron stars
or black holes in the dependence of the concrete initial mass value, see \citet{HFWLH}.
The masses of the stellar remnants can be evaluated with the data from \citet{HFWLH}.

7. The uncertainties of the estimates are evaluated by variation
of the exponents of the mass spectrum within the ranges
[-1.8;-0.8] and [-2.6;-2.0], and by taking
into account the uncertainties of the stellar masses from Table 9,
and the uncertainty of the cluster star number from Table 11.

8. The presence of unresolved binary stars can be taken into account
following \citet{KB} and supposing, for example, the same binary fraction
as in the Praesepe cluster (0.35). In that case
the coefficient 1.35 should be applied to the mass estimate.

Applying these steps to NGC 1502 gives the estimate of NGC 1502
total mass between approximately 1760
and 3900 solar masses. The uncertainty of this estimate is very large.
Moreover, the fraction of the unresolved binary stars
can vary in the range from 0.3 to 0.5 \citep{bin_fr}.
Due to large uncertainty, this procedure was not applied to
the sample clusters; it was preferred to use the lower mass estimates
listed in Table 12 for all sample clusters (including NGC 1502).

With this lower estimates of the sample cluster masses,
the lower estimates of the cluster tidal radii in the Galactic
gravitational field were calculated. The model of Galactic gravitational
potential $\Phi$ was used from \citet{KO}.
The following formula was used for the tidal radii estimate \citep{King}:

\begin{equation}
\label{tidalrad}
R_{\rm t}=\left( \frac{GM}{4A(A-B)}\right)^{1/3}=\left( -\frac{GM}{\alpha_1}\right)^{1/3} \; \mbox{.}
\end{equation}

Here G is the gravitational constant, $G=0.004535$ in the unit system
1 pc for distance; 1 $\rm M_{\odot}$ (one solar mass) for mass;
and 1~Myr for time, as adopted in the present work. M is the cluster mass;
A and B are Oort's constants for the cluster Galactocentric distance $R_{\rm cl}$; $\alpha_1$ is
the parameter describing the Galactic potential at the current Galactocentric
distance of the cluster (introduced by \citet{Chandrasekhar}):

\begin{equation}
\label{alpha1}
\alpha_1= R\left.\left(\frac{1}{R}\frac{\partial \Phi}{\partial R}- \frac{\partial^2\Phi}{\partial R^2}\right)\right|_{R=R_{\rm cl}} \; \mbox{,}
\end{equation}

\noindent where $R$ is the distance from the Galactic centre and $R_{\rm cl}$
is the cluster distance from the Galactic centre.

\begin{equation}
\label{galdist}
R_{\rm cl}=\sqrt{R_0^2+d^2\cos^2b-2R_0d\cos l\cos b} \; \mbox{,}
\end{equation}

\noindent where $R_0$ is the Solar distance from the Galactic centre
($R_0=8200$ parsecs value was taken here, see, for example, \citet{Nik}
and \citet{HouHan}), $l$ and $b$ are the galactic
coordinates of the cluster, and $d$ is the cluster distance from the Sun.
With the \citet{KO} model,

\begin{eqnarray}
\label{alpha11}
\alpha_1 & = & -2\Phi_0\left(\frac{R_{\rm cl}}{R_{\rm a}^2}\right)^2\frac{1+3e}{e^3(1+e)^3} \; \mbox{, and} \\
e & = & \sqrt{1+\left(\frac{R_{\rm cl}}{R_{\rm a}}\right)^2} \; \mbox{,}\nonumber
\end{eqnarray}

\noindent where $R_{\rm a}=2000$ pc, and $\Phi_0=1.841\cdot10^5 \; \rm pc^2/Myr^2$.

The Galactic potential model of \citet{KO} was chosen on the following considerations.
In order to derive the open cluster tidal radii, the model of Galactic potential
is needed, that well describes the Galactic potential in the Solar vicinity
in the Galaxy, because all the sample clusters are close to the Sun ($d<2.36$ kpc).
The compatibility of the Oort constants A and B derived from the model
and modern data on the A and B can be a criterion. \citet{BoBa} determined $A=16.49\pm0.60 \; \rm{km/s/kpc}$ and
$B=-12.37\pm1.12 \; \rm{km/s/kpc}$ with the study of high precision data on the 73 maser
sources. These values give $4A(A-B)\simeq1900 \; \rm{km^2/s^2/kpc^2}$. The constants A and B derived
from the \citet{KO} model with $R_0=8200$ parsecs are $A=17.08 \; \rm{km/s/kpc}$ and
$B=-10.58 \; \rm{km/s/kpc}$. These values give $4A(A-B)\simeq1890 \; \rm{km^2/s^2/kpc^2}$, that is
very close to the value from \citet{BoBa}.

The Solar Galactocentric distance of $R_0=8200$ parsecs is the reasonable value,
compatible with the modern data, see the reviews in \citet{Nik} and \citet{HouHan}.

The modern models of the Galactic potential are aimed at the determination of the
Galactic extended dark halo parameters, see, for example, \citet{Bonaca}.
The perturbations are added to the potential, that are connected with the presence of the bar
and the spiral arms, see the review in \citet{Pettitt}. But the Galactic
potential model of \citet{KO} is relatively simple, and gives the adequate values of the Oort constants
in the Solar vicinity, and it is sufficient for the present work.

The lower estimates of the sample cluster tidal radii are listed
in the third column of Table 12. The uncertainty of this
estimate was obtained taking into account the uncertainty
of the cluster mass estimate, the uncertainty of the cluster distance from the Sun,
and a 10\% uncertainty of the $R_0$ value.

The fourth column of Table 12 contains a maximum visual estimate
of the cluster radius for all magnitude intervals. The fifth
column of Table 12 contains the maximum corona radius for all
magnitude intervals, obtained by the cluster surface density
profile approximation with the combined model.
It is seen that NGC 6819 and NGC 6939 extend well beyond
their tidal surfaces. This fact is unlikely to be changed due to
the unknown low-mass tail of stellar content
in these clusters and to unresolved binaries, because Eq.(\ref{tidalrad}) contains
the cluster mass to the $1/3$ power. Then an increase of the
cluster mass by two times will lead to a tidal radius increase
by only a 1.26 factor. The large extension of these clusters
can be explained by their non-stationarity: the rapid expansion of the cluster and
the stabilizing ejections of the cluster stars into
galactic field (see \citet{D82,D05,D11}).

The young and intermediate-age clusters can be subjected to the
influence of additional gravitational action from the
nearest gas-star complex with concomitant movement relative to the cluster
(that is the gas-star complex where the cluster has been formed).
This action leads to a decrease
in the cluster tidal radius of a factor of 1.5-2.5 \citep{D90}.
Taking into account this possibility, it can be so explained
why young and intermediate-age clusters from our sample
show the same evidence of non-stationary processes (see Fig.11)
as old clusters NGC 6819 and NGC 6939, which extend over
their tidal surfaces.

\section{Conclusions}
The purpose of the present study was to show the efficiency of kernel
estimation of surface and spatial density profiles
of open star clusters and their N-body models, especially in the outer
cluster region, and to demonstrate the necessity of taking into account
the corona component of the open cluster when choosing the
model for the surface density profile approximation.

The following general results were obtained in the present research.

1. The formulae for kernel estimates of spatial density profiles
of star clusters were obtained, for the cases when stellar
spatial coordinates (x,y,z) are known. Spatial density profiles
for N-body models of open cluster coronae were derived as
examples. The result of
\citet{DPS} was confirmed concerning the formation of
quasi-equilibrium density distribution in the open cluster
coronae up to distances of three tidal radii from the
cluster centre.

2. Surface density profiles were derived for seven open clusters
for different limiting magnitudes using the data of 2MASS. The optimal
kernel halfwidth value was selected following \citet{MT}, it was
the value, that gave the smoothest curve that closely followed the mean
trend defined by curves computed with much smaller kernel halfwidth.
The surface density of the stellar background and cluster radii were
estimated by the surface density profile. It was shown that the cluster
radius estimate is hardly dependent on the kernel halfwidth value,
when it is less or equal to the optimal one. The comparison with other
investigations shows that data on open cluster sizes are often underestimated.
The result of \citet{AK} was confirmed about the presence of an extended corona
in the open cluster NGC 6939.

3. The surface density profiles of the sample clusters show
evidence of mass segregation and irregularities in the outer
parts of clusters which can be interpreted as evidence of
non-stationary processes in the clusters.

4. The surface density profiles of the sample clusters were approximated
by the King function and by the combined model; that is, a combination of
the King function for the cluster core and the uniform sphere
for representation of the cluster corona. It is shown that
the combined model describes surface density profiles of the sample clusters
much better than the King model alone. This is especially
well seen when the cluster star numbers, obtained by
integration of the surface density profiles from the kernel estimates
and its models, are compared.

5. The lower estimates of the sample cluster masses and tidal radii
in the Galactic gravitational field were obtained. It is shown that
open clusters NGC 6819 and NGC 6939 extend beyond their
tidal radii. This can be explained by their non-stationarity: by rapid
expansion of these clusters and by the stabilizing ejections of the cluster
stars into the galactic field.

\section*{Acknowledgments}
The author is very grateful to Prof. D.Merritt for
introducing him to the kernel estimator method, and
to Prof. V.M.Danilov for helpful discussions.
The author also acknowledges the assistance of
Dr. T.P.Rasskazova and Mr. Ian Miller (Dept. Foreign Languages,
INS, UrFU) in the preparation of this article.
This work was partially supported by the Ministry of Education
and Science of the Russian Federation (state contract
No. 3.1781.2014/K, registration number 01201465056).
The travel to the conference was supported
by Act 211 Government of the Russian Federation, agreement No. 02.A03.21.0006.
This publication makes use of data products from the Two Micron All Sky Survey, which
is a joint project of the University of Massachusetts and the Infrared Processing
and Analysis Center, California Institute of Technology, funded by the
National Aeronautics and Space Administration and the National Science Foundation.

\end{document}